\documentclass[apjl]{emulateapj}

\usepackage{rotating}
\usepackage{afterpage}
\usepackage{txfonts}
\usepackage{graphicx}
\usepackage{rotating}
\usepackage{afterpage}
\usepackage[]{natbib}
\bibpunct{(}{)}{;}{a}{}{,}

\newcommand{\um}{$\mu$m}
\newcommand{\brgamma}{Br$\gamma$}
\newcommand{\kms}{km\thinspace s$^{-1}$}

\def\arcsec{\hbox{$^{\prime\prime}$}}
\def\utw{\smash{\rlap{\lower5pt\hbox{$\sim$}}}}
\def\udtw{\smash{\rlap{\lower6pt\hbox{$\approx$}}}}

\def\Mdot{\hbox{$\dot {M}$}}

\def\Rsun{\hbox{\it R$_\odot$}}

\def\Lsun{\hbox{\it L$_\odot$}}

\def\Teff{\hbox{\it T$_{\rm eff}$}}

\def\Msun{\hbox{\it M$_\odot$}}

\def\Msunyr{\hbox{\it M$_\odot\,$yr$^{-1}$}}

\def\kpc{\hbox{kpc}}
\def\pc{\hbox{pc}}
\def\Jy{\hbox{Jy}}
\def\Teff{\hbox{\it T$_{\rm eff}$}}
\def\Vinf{\hbox{$v_\infty$}}

\newcommand{\Aks}{{\it A$_{\rm K_{\rm s}}$}}

\def\simgr{\mathrel{\hbox{\rlap{\hbox{\lower4pt\hbox{$\sim$}}}\hbox{$>$}}}}

\def\vlsr{\hbox{\it V$_{\rm LSR}$}}
\def\iras{\hbox{IRAS~16115$-$5044}}

\DeclareRobustCommand{\ion}[2]{%
\relax\ifmmode
\ifx\testbx\f@series
{\mathbf{#1\,\mathsc{#2}}}\else
{\mathrm{#1\,\mathsc{#2}}}\fi
\else\textup{#1\,{\mdseries\textsc{#2}}}%
\fi}

\shorttitle{A New Candidate LBV}
   
\shortauthors{Figer et al.}

\begin{document}

\title{A new candidate Luminous Blue Variable}

\author{Donald F. Figer\altaffilmark{1},
	Francisco Najarro\altaffilmark{2},
	Maria Messineo\altaffilmark{3}, 
        J. Simon Clark\altaffilmark{4},
        Karl M. Menten\altaffilmark{5}
        }

\email{figer@cfd.rit.edu}
\altaffiltext{1}{Center for Detectors, Rochester Institute of Technology, 54 Memorial Drive, Rochester, NY 14623, USA}
\altaffiltext{2}{Centro de Astrobiología (CSIC/INTA), ctra. de Ajalvir km. 4, 28850 Torrejón de Ardoz, Madrid, Spain}
\altaffiltext{3}{Key Laboratory for Researches in Galaxies and Cosmology, University of Science and Technology of China, Chinese Academy of Sciences, Hefei, Anhui, 230026, China}
\altaffiltext{4}{Department of Physics and Astronomy, The Open University, Walton Hall, Milton Keynes, MK7 6AA, UK}
\altaffiltext{5}{Max-Planck-Institut f\"ur Radioastronomie, Auf dem H\"ugel 69, D-53121 Bonn, Germany}

\begin{abstract} 
We identify \iras, which was previously classified as a protoplanetary nebula (PPN), as a candidate luminous blue variable (LBV). 
The star has high luminosity ($\simgr$$10^{5.75}$~\Lsun), ensuring supergiant status, has a temperature similar to LBVs, is photometrically and spectroscopically variable, and is surrounded by warm dust.
Its near-infrared spectrum shows the presence of several lines of \ion{H}{I}, \ion{He}{I}, \ion{Fe}{II}, \ion{Fe}{[II]}, \ion{Mg}{II}, and \ion{Na}{I} with shapes ranging from pure absorption and P~Cygni profiles to full emission. 
These characteristics are often observed together in the relatively rare LBV class of stars, of which only $\approx$20 are known in the Galaxy. 
The key to the new classification is the fact that we compute a new distance and extinction that yields a luminosity significantly in excess of those for post-AGB PPNe, for which the initial masses are $<$8~\Msun. Assuming single star evolution, we estimate an initial mass of $\approx$40~\Msun.
\end{abstract}


\section{Introduction} 
Luminous blue variables (LBVs) have a distinct set of observed characteristics \citep{conti84}. 
They have luminosities of supergiants, temperatures above 10~kK, non-periodic variability, and evidence of eruptions vis a vis circumstellar ejecta \citep[e.g.][]{clark05}. 
They are inferred to be post-main sequence descendants of massive stars \citep{meynet11}. 
Their spectrophotometric variations are on the order of 1--2 mag over timescales of years at roughly constant bolometric luminosity.
During giant eruptions, their brightness changes by up to 3 mag \citep[e.g.][]{vangenderen97}.
They are often surrounded by ionized gas and warm dust, both evidence of past eruptions \citep{clark05}. 
The archetypal example, LBV $\eta$ Car, is surrounded by the Homunculus, material inferred to have been ejected by the star during the great eruption starting in 1837 \citep{smith06}. 
These eruptions carry with them an extraordinary amount of material, such as the $\approx$10-20~\Msun\ of material surrounding the $\eta$~Car \citep{smith07}.
The effective mass-loss rates during these eruptions is extraordinarily high, such as up to $\sim$1~\Msun\ yr$^{-1}$ for $\eta$~Car during its great eruption. 
Their evolutionary paths in the Hertzsprung-Russel diagram are uncertain, due in part to poor statistics, but they are clearly near, or sometimes even above, the Humphreys-Davidson limit \citep{humphreys94}.
There are approximately 20 known LBVs in the Galaxy and a similar number of candidates \citep{smith19}.
LBVs may be progenitors of at least some supernovae \citep[c.f.,][]{smith11,burgasser12,dessart15}, and \citet{allan20} argue that an LBV directly collapsed to a black hole.

In this paper, we demonstrate that \iras\ (G332.2843-00.0002) has all the characteristics of LBVs, confirming the claim in \citet{messineo20}.
Located in a complex region containing many massive stars and compact objects \citep{messineo20}, the star was previously classified as a protoplanetary nebula (PPN) \citep{weldrake03}, an evolved post-AGB star having initial mass less than 8~\Msun.
We argue that it is more appropriately identified as an LBV, similar to how He 3-519 and the Pistol star were reclassified from PPNe to LBVs \citep{figer98}.
In this Letter, we review available observations, present newly-reduced spectra, and reinterpret the nature of this star.
In Sect. \ref{secdata}, we present photometry from the literature and newly-reduced spectroscopy. 
In Sect. \ref{secdistance}, we estimate the distance. 
In Section \ref{secneb}, we describe and characterize the dusty nebula. 
In Section \ref{secext}, we estimate the extinction. 
In Section \ref{seclum}, we estimate the stellar luminosity.
In Section \ref{remark}, we compare the star to other LBVs. 

\section{Data}
\label{secdata}

\subsection{Spectra from the Literature}

\citet{suarez06} obtained an optical spectrum of the star in June 1990 using the 1.5~m ESO telescope.
They were searching for planetary nebulae (PNe), and categorized IRAS~16115$-$5044 as a ``young'' star.
The spectrum has an emission line (see their Figure D.1.), but we suspect that the object might have been misidentified. 
Note that the object is expected to have V$\gtrsim$21, given that it has G=14.6 from Gaia and G-J$\sim$7, and such an object would be a challenge to observe with a 1.5~m telescope.

\citet{Oudmaijer95} published a K-band spectrum with relatively low signal-to-noise ratio (S/N) of the object. 
It appeared featureless in the narrow wavelength range they covered (2.2--2.4~\micron), except perhaps for a \ion{Mg}{II} line near 2.40~\micron. Several years later, 
\citet{weldrake03} observed \iras\ at infrared wavelengths, assigning a spectral type of B4Ie and an wind outflow velocity of 300~\kms. 
They identified emission lines in the hydrogen Paschen and Bracket series, as well as from \ion{Fe}{II} and 2.089~\um, \ion{Mg}{II} at 2.138 and 2.144~\um, and \ion{Na}{I} at 2.206 and 2.209~\um. 

\subsection{New Spectra}
\label{secspectra}

\begin{figure*}
\begin{center}
\resizebox{1.\hsize}{!}{\includegraphics[angle=0]{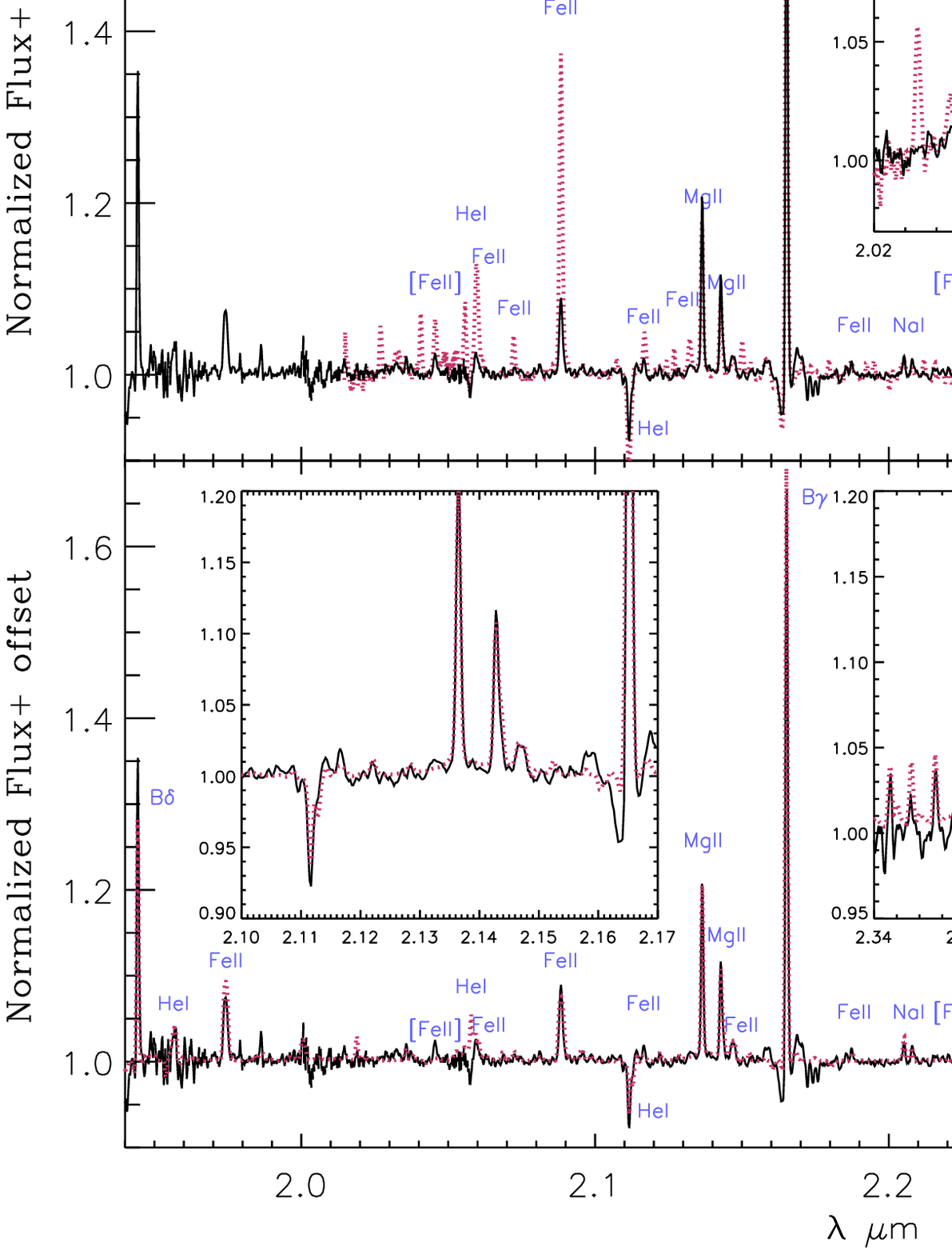}}
\caption{\label{kbandsinfoni} The upper panel compares the 2013 (dashed red) and 2016 (solid black) spectra of IRAS~16115$-$5044 taken with SINFONI, as obtained from the ESO Science Archive Facility. The lower panel compares the 2016 data with a CMFGEN model fit (dashed red). The insets show close up sections of the spectra. Line identifications are from \citet{figer98}. See \citet{najarro09} and \citet{voors00} for similar spectra of qF362 and and G79.29+0.46, respectively. }
\end{center}
\end{figure*}

Figure \ref{kbandsinfoni} shows two observed spectra (solid-black), corrected for telluric absorption and emission, obtained with SINFONI on the ESO Very Large Telescope (VLT), together with a model fit (dashed red line, see Sect.\ref{seclum}).
The spectrum in the upper panel was observed in 2013 with a 0$\farcs$25 plate scale (Program ID: 091.D-0376(A)), whereas that in the lower panel was observed in 2016 with a 0$\farcs$025 plate scale (Program ID: 097.D-0033(A)) .
The spectra were reduced using the ESO pipeline with wavelength calibration set by OH lines for the 2013 data set.  
These lines were not visible in the 2016 data, so we used arc lamp lines in data taken during the daytime. 
After the wavelength calibration, we measured the centroids of the observed emission lines and estimate
the S/N which varies from $\sim$70 in the telluric polluted regions to $\sim$250, as measured near 2.10~\micron. 
The spectra display strong Br-$\gamma$ emission, with a weak P~Cyg absorption dip and the corresponding \ion{He}{I} hydrogenic components in absorption. 
\ion{H}{I} lines in the Pfund series up to Pf$_{30}$ are clearly seen in emission longward of 2.3~\micron, with a P~Cygni shape present in higher-resolution and S/N spectrum from 2016 which decreases from  Pf$_{19}$ till Pf$_{28}$.
The 2.112/2.113~\micron\ \ion{He}{I} doublet lines are in absorption, and the 2.058~\micron\ \ion{He}{I} line is in emission with a potential P~Cygni profile.
This line is blended with the \ion{Fe}{II} 2.060~\micron.
Two \ion{Mg}{II} doublets in emission are seen near 2.14 and 2.41~\micron. 
\ion{Fe}{II} and \ion{Fe}{[II]} emission lines are seen throughout the spectrum. 
The \ion{Na}{I} doublet is detected near 2.21~\micron.
Many lines are present in similar strengths in both spectra. The iron and helium lines are clearly stronger in the 2013 spectrum (upper panel). 
The spectrum looks similar to that in \citet{weldrake03}, although the \ion{Fe}{II} line strengths in the 2016 spectrum provide a best match. 

The spectra are nearly identical to those for qF362 \citep{najarro09} and G79.29+0.46 \citep{voors00}. 
They are also similar to those of other LBVs, e.g., AG Car, the Pistol star (12~kK),  G24.73+0.69 (12~kK), G26.47+0.02 (17~kK), and G0.120-0.048  \citep{morris96,najarro09, clark03, mauerhan10,clark18}.
The \ion{He}{I} lines suggest a temperature $>$10.5~kK, and the \ion{Na}{I} lines suggest a temperature $<$13~kK, which taken together suggests a spectral type of B5-8 at the time of the observation \citep[see Figure 8 in][]{messineo11}.
The widths from the \ion{Fe}{II} 1.974 and 2.089~\micron\ emission lines which are formed in the mid-outer wind and hence provide a reliable estimate of \Vinf\ are $\approx$350~\kms, suggesting a wind speed ($\sim 175$~\kms) that is typical for LBVs \citep{smith14}.

IRAS~16115$-$5044 has a variable spectrum, particularly in the \ion{Fe}{II}, [\ion{Fe}{II}], and \ion{He}{I} lines, although the hydrogen  and \ion{Mg}{II} lines are more constant.
The average equivalent width of the \brgamma\ line is 172~\AA$\pm$3.7\%, and those of the \ion{Mg}{II} lines are 26~\AA$\pm$3.0\% and 17~\AA$\pm$7.7\%, where the percentage error represents the difference between the two measurements and the mean. 
The equivalent widths for some of the iron lines vary by up to 400\%. 
The shape of the blend of the apparent \brgamma\ P~Cygni absorption line and the \ion{He}{I} photospheric line also appears to change between the two observations. 

\subsection{Photometry}
\label{secphot}

The data in Table \ref{tablephot} were taken at many different times and with very different beam widths. The circumstellar emission is clearly included for some of them, e.g., in the IRAS and Akari data.

\begin{table*}
\caption{ \label{tablephot} Photometric measurements of \iras.}
{\tiny
\begin{center}
\begin{tabular}{llrrrll}
\hline
\hline
Survey      & band   &$\lambda$& Flux   & mag  & Name & Reference   \\
         &      & [\um]  & [\Jy]   & [mag] &\\
\hline
     GSC2.2  &    R &  0.70 &  0.00 & 17.15 &    S230213364722 &  \citet{lasker08}  \\
     USNOB1  &   R2 &  0.70 &  0.00 & 17.79 &     0391-0542333 &   \citet{monet03}  \\
     USNOB1  &  Imag &  0.90 &  0.01 & 13.47 &     0391-0542333 &   \citet{monet03}  \\
     DENIS  &    I &  0.79 &  0.01 & 13.05 &   J161517.9-505219 & \citet{epchtein94}  \\
     2MASS  &    J &  1.23 &  1.25 &  7.76 &   16151795-5052197 &   \citet{cutri03}  \\
     DENIS  &    J &  1.22 &  1.38 &  7.67 &   J161517.9-505219 & \citet{epchtein94}  \\
     2MASS  &    H &  1.63 &  3.58 &  6.13 &   16151795-5052197 &   \citet{cutri03}  \\
     2MASS  &    K &  2.15 &  5.80 &  5.14 &   16151795-5052197 &   \citet{cutri03}  \\
    GLIMPSE  &   4.5 &  4.50 &  5.38 &  3.81 &  G332.2843-00.0008 &  \citet{spitzer09}  \\
    GLIMPSE  &   5.8 &  5.80 &  3.83 &  3.69 &  G332.2843-00.0008 &  \citet{spitzer09}  \\
    GLIMPSE  &   8.0 &  8.00 &  3.02 &  3.32 &  G332.2843-00.0008 &  \citet{spitzer09}  \\
      WISE  &   W3 &  11.56 &  3.31 &  2.36 & J161518.17-505220.4 &  \citet{wright10b}  \\
    Mipsgal  &   F24 &  23.70 & 106.00 & -2.85 & MGE332.2843-00.0002 & \citet{gutermuth15}  \\
      IRAS  &   F12 &  12.00 & 10.40 &  1.09 &    IRAS16115$-$5044 &    \citet{iras}  \\
      IRAS  &   F25 &  25.00 & 185.00 & -3.60 &    IRAS16115$-$5044 &    \citet{iras}  \\
      IRAS  &   F60 &  60.00 & 500.00 & -6.56 &    IRAS16115$-$5044 &    \citet{iras}  \\
      IRAS  &  F100 & 100.00 & 255.00 & -6.93 &    IRAS16115$-$5044 &    \citet{iras}  \\
     Akari  &   S9 &  9.00 &  3.26 &  3.09 &    1615181-505218 &\citet{ishihara10},\  \\
     Akari  &   S65 &  65.00 & 337.80 & -6.36 &    1615181-505218 &\citet{ishihara10},\  \\
     Akari  &   S90 &  90.00 & 132.10 & -5.81 &    1615181-505218 &\citet{ishihara10},\  \\
     Akari  &  S140 & 140.00 & 81.99 & -6.59 &    1615181-505218 &\citet{ishihara10},\  \\
     Akari  &  S160 & 160.00 & 56.25 & -6.45 &    1615181-505218 &\citet{ishihara10},\  \\
\hline
\end{tabular}
\end{center}
}
\end{table*}

IRAS~16115$-$5044 is in the Infrared Astronomical Satellite Point Source Catalogue (IRAS PSC) \citep{iras}, with a flux density of 10.4~\Jy\ at 12~\um, 185~\Jy\ at 25~\um, 500~\Jy\ at 60~\um, and 255 Jy at 100~\um. 
The IRAS colors are [12]-[25]=4.68 mag  and [60]-[100]=0.37 mag.
In the van der Veen/Habing diagram \citep[Figure 5b in][]{vanderveen88}, the [60]-[100] color falls in region V, though the [12]-[25] is redder than 2 mag.
Region V is the region of PNe and non-variable stars with cold envelopes.

\subsection{Photometric variations}

The source was analyzed for photometric variability with data from the Diffuse Infrared Background Experiment (DIRBE) instrument on the Cosmic Background Explorer (COBE) at 1.25~\um, 2.2~\um, 3.5~\um, and 4.9~\um\ and reported as a non-variable star \citep{price10}.
The flux density increased over the 3.6 year time period of the observations.
In the $J$-band, we measure a linear flux increase from 80 counts on 1990 February 1 to 100 counts on 1993 April 8 (0.24 mag). 
The standard deviation is $0.23$ mag, with similar variations in the other bands, consistent with what is observed for some other LBVs.
For example, from 1985 to 1992, AG Car showed a steady small flux increase with variations of 0.1 mag \citep[e.g.][]{vangenderen97}.


\section{Distance}
\label{secdistance}

We measured an LSR velocity, \vlsr, of $-$58.6$\pm$3.8~\kms\ using the \ion{Mg}{II} lines in the 2013 and 2016 SINFONI data for which we were able to measure the locations of telluric OH lines in order to set the wavelength scale. 
This \vlsr\ was confirmed, within the uncertainties, by cross-correlating our spectroscopic model (see Sect.\ref{seclum}) with the 2016 spectra, denoting that the shapes of the \ion{Mg}{II} lines are barely affected by the stellar wind.
Note that we validated our wavelength calibration technique by applying it to SINFONI data sets for GG~Car and MCW~137, both early type supergiants with significant winds and near-infrared spectral morphology similar to that of \iras. 
In both cases, we estimated \vlsr\ and inferred distances consistent with values in the literature.
Using the A5 model of \citet{reid14}, we estimate a distance of 3.68$\pm$0.35~\kpc,
similar to that of nearby supernova remnant RCW103 and PSR~J1616-5017 \citep{messineo20}. 
The error is the quadrature sum of the error from the model plus the implied distance error based on the error in the velocity measurement.

The high \Aks\ determined in Section \ref{secneb} is consistent with the kinematic distance \citet{messineo20}.
Radio wavelength absorption has been observed in the 1612, 1665 and 1667 GHz hyperfine structure lines of the OH molecule toward the position of of \iras\ by \citet{telintel96}. Absorption along its line of sight was detected in all three lines (the 1720 MHz lines was not covered). The LSR velocities of the OH absorption features vary from line to line with common intervals and in total cover LSR velocities from $-100$ to $-35$~\kms. \citet{urquhart07} published a spectrum of the $^{13}$CO $J=1-0$ lines of this source, taken as part of the Red MSX Source (RMS) survey effort. This line shows various emission features with LSR velocities between $-109$ and $-45$~\kms, similar to the velocities covered by the OH lines.
Given the ubiquity of OH and $^{13}$CO in the interstellar medium and the relatively large beam widths used in the radio studies, it is impossible to establish a direct relation of the molecular gas with \iras. 
We note that the depth of the OH absorption, up to $-$0.7~Jy, indicates that the absorption must be against extended Galactic background radiation, as no compact radio source has been detected toward \iras. 
\citet{messineo20} discuss the overlapping clouds along this line of sight. In summary, we find that the extinction and radio features are consistent with the kinematic distance inferred from the infrared spectra.

From the considerations in this section, we adopt a distance of 3.68~\kpc\ for \iras, placing it in the Scutum-Crux spiral arm of the Galaxy \citep[see Figure 6 in ][]{messineo20}.

Note that we inspected the GAIA DR2 database, finding a parallax of $\varpi$=0.69$\pm$0.19~mas~yr$^{-1}$, giving a distance range of 1.1 to 2~\kpc. 
We give this distance little weight, as comparison of distances inferred from GAIA data with spectrophotometric distances for OB stars implies deviations up to 50\% for distances $>2$~\kpc\ \citep{shull19}.

\section{Warm dust}
\label{secneb}

\begin{figure*}[t]
\begin{center}
\resizebox{0.45\hsize}{!}{\includegraphics[angle=0]{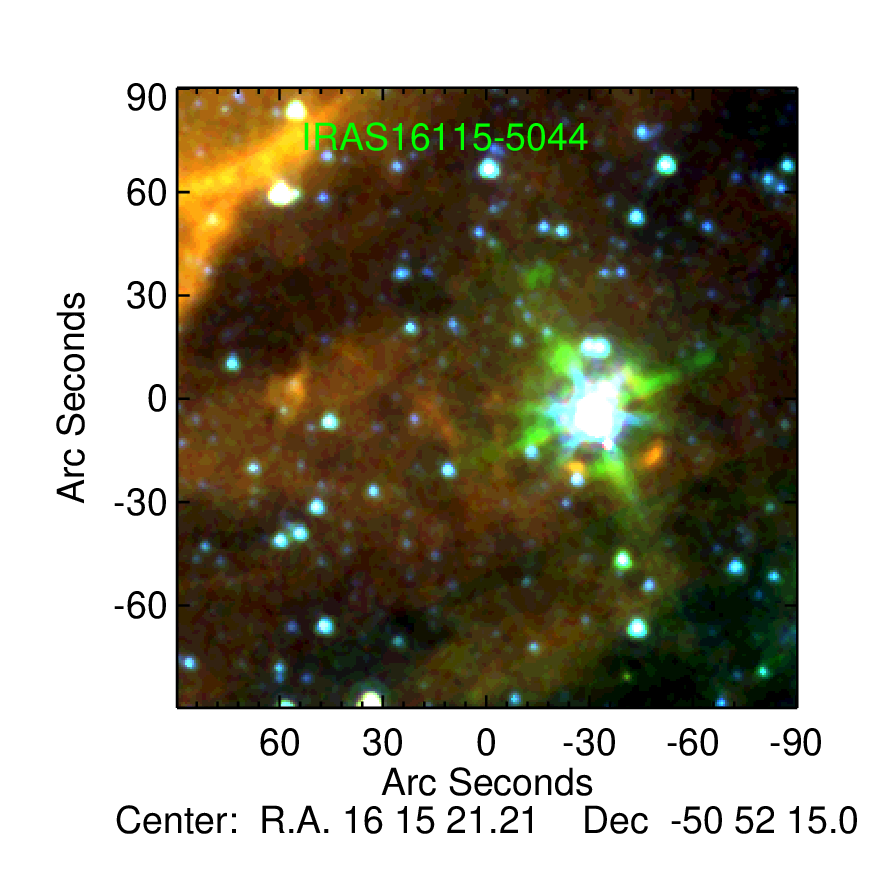}}
\resizebox{0.45\hsize}{!}{\includegraphics[angle=0]{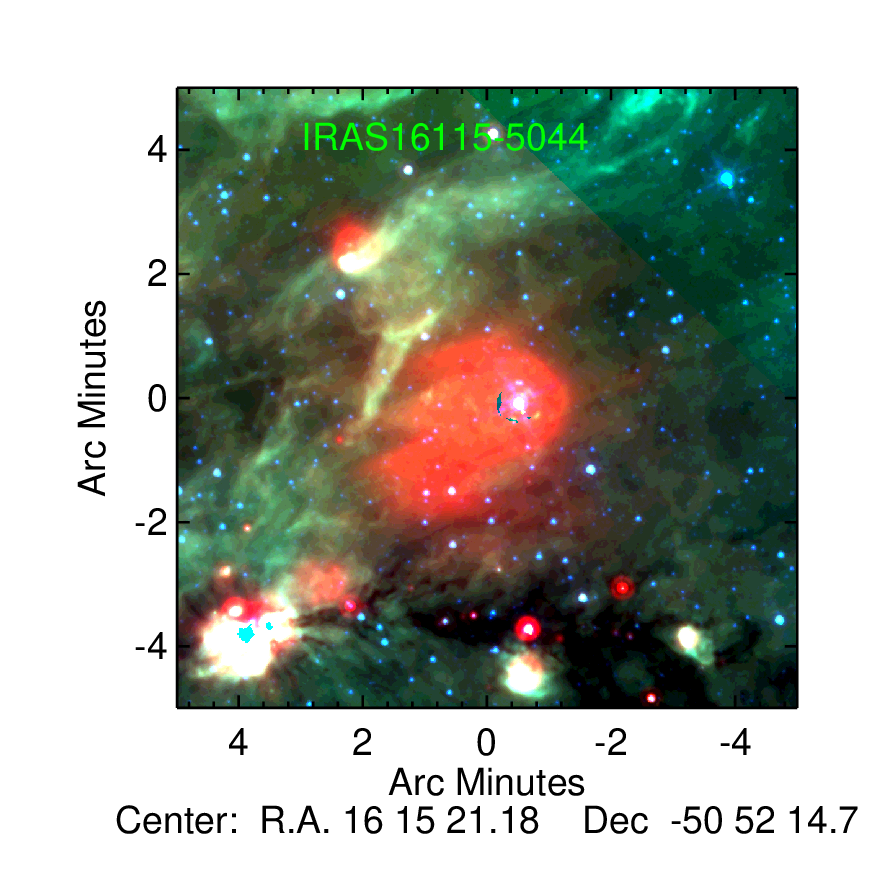}}
\caption{ \label{figmips} ({\it left})
Composite GLIMPSE image at 3.6~\um\ (blue), 5.8~\um\ (green), and 8.0~\um\ (red).
({\it right})
Composite GLIMPSE/MIPSGAL image at 5.8~\um\ (blue), 8.0~\um\ (green), and 24~\um\ (red).}
\end{center}
\end{figure*}

Two composite images of the region from the GLIMPSE and MIPSGAL surveys are shown in Figure \ref{figmips}.
IRAS~16115$-$5044 is a bright 24~\um\ source (106~\Jy) and is listed in the catalog of \citet{mizuno10} as an extended source with a diameter of 47\arcsec, corresponding to physical size of 0.8~\pc\ for a distance of 3.6~\kpc, sizes that are typical for LBVs \citep{nota95}.
The object is below the detection threshold of \citet{vandeSteene93} at radio wavelengths, suggesting that the star is not ionizing the nearby dust.
This precludes the typical straightforward determination of the nebular mass since any determination from the properties of the dusty component would require a somewhat arbitrary dust:gas ratio to be adopted.
We also note the apparent asymmetric nature of the nebula at 24~\micron\ (Figure \ref{figmips}, right).


\section{Interstellar extinction}
\label{secext}

IRAS~16115$-$5044 was classified as a PPN because the inferred luminosity was consistent with those of the central stars in PNe (4,000-6,000 \Lsun), but this was due to an incorrect estimate of interstellar extinction.

\begin{figure}[h]
\begin{center}
\resizebox{1\hsize}{!}{\includegraphics[angle=0]{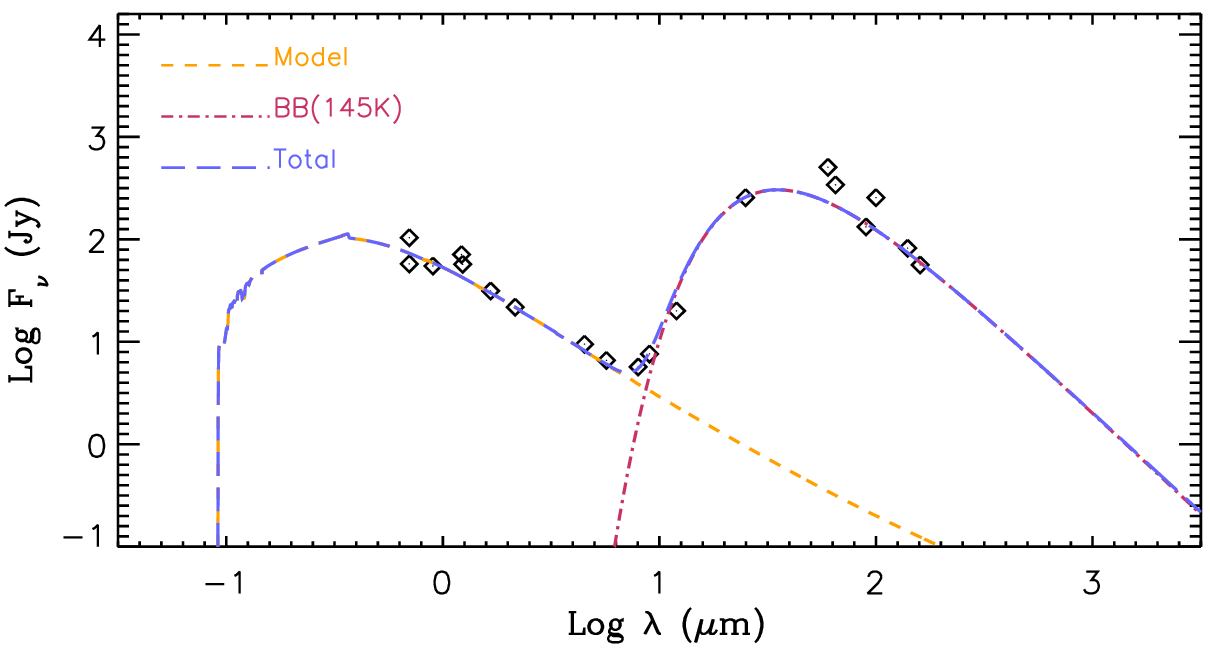}}
\end{center}
\caption{ \label{multi} Dereddened photometry of IRAS~16115$-$5044 (black diamonds), total flux (dashed long blue) from the model SED (dashed orange) plus a T=145~K blackbody (dashed dotted red).}
\end{figure}

Using the spectral energy distribution (SED) from the spectroscopic model, we estimate that the intrinsic near-infrared colors are close to zero ($H-K$=0.08 and
$J-H$=0.13). 
Applying curve number three from \citet{messineo05} for the extinction law to the observed color excesses, we estimate \Aks=1.4. 
Figure \ref{multi} shows a model fit to the dereddened photometry. 
It consists of the stellar model SED plus the dust contribution provided by a black-body with a temperature of 145~K.

\section{Luminosity and mass}
\label{seclum}

We modeled the SED and K-band spectrum with CMFGEN \citep{hillier98} in a process similar to that described in \citet{najarro09}, finding $log(L/{L_\odot})=5.75$, \Teff=11.0~kK ($\tau_{Ross}$=2/3, R=205~\Rsun), 
V$_{inf}$=170\kms, and \Mdot=4.75(10$^{-6}$)~\Msunyr\ with a moderate clumping (f$_{cl} \sim 0.08$) initiating close to the base of the wind. 
As in \citet{najarro09}, we are able to break the He/H degeneracy and obtain He/H=0.40 by number. 
Figure \ref{kbandsinfoni} displays the excellent fit of our spectroscopic model to the observations for the 2016 data set. 

We refrained from modelling the 2013 data, due to its lower S/N compared to the 2016 data. 
However, we can qualitatively say that the 2013 spectrum corresponds to a slightly higher temperature phase reflected by the \ion{He}{I} components, the increased strength of the \ion{Fe}{II} lines, and the weakened \ion{Na}{I} lines.  

\begin{figure}
\begin{center}
\resizebox{1\hsize}{!}{\includegraphics[angle=0]{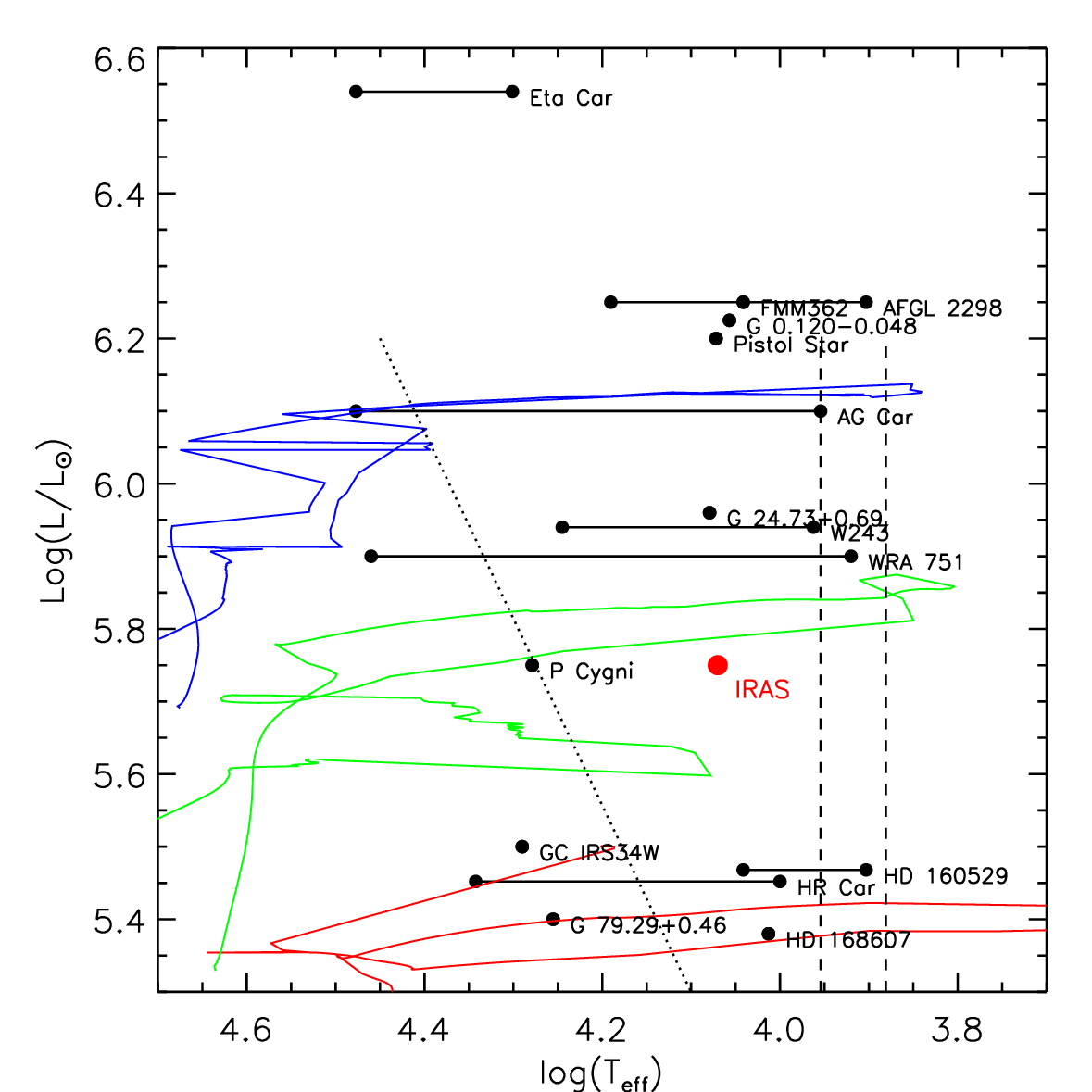}}
\end{center}
\caption{ \label{lbvhrd} HR diagram for \iras\ and other LBVs.
Data for other objects are from 
Eta Car \citep{humphreys94}, 
P Cyg \citep{najarro97}, 
HD~168607 \citep{leitherer84},
AG~Car \citep{groh09},
HR~Car \citep{boffin16},
HD~160529 \citep{humphreys94},
Wra~751 \citep{nakau13},
qF362 \citep{najarro09},
AFGL~2298 \citep{clark09},
G24.73+0.69 \citep{clark03},
W243 \citep{ritchie09},
GCIRS34W \citep{martins07},
G0.120-0.048 \citep{mauerhan10},
Pistol star \citep{figer98,najarro09}, 
and G79.29+0.46 \citep{voors00}.
Estimates of the minimum and maximum temperatures are connected with dashed lines, when available.
The two long-dashed vertical lines enclose the outburst region of LBVs as shown in \citet{smith04}.
The diagonal dotted line marks the hot edge of the LBV minimum strip \citep{clark05,smith04}. 
Three rotating stellar tracks of 25, 40, and 60~\Msun\ are plotted with red, green, and blue curves, respectively \citep{ekstrom12} .
We set the values for G0.120-0.048 as intermediate between those of the Pistol star and FMM362, as the K-band spectra are all similar \citep{figer98,geballe00,mauerhan10}
}
\end{figure}

Figure \ref{lbvhrd} plots the location of the object on the HR diagram, along with data points for LBVs and evolutionary tracks for rotating massive stars \citep{ekstrom12}.
From the location of \iras\ in the HR diagram, it appears that the initial mass for the star is $\approx$40~\Msun, assuming that the object is a single star.
While it is possible that it could be a multiple star system, we note that there are no indications of multiplicity, or binary star evolution, in the spectra.

\section{Conclusions}
\label{remark} 

We revise the classification of IRAS~16115$-$5044 from a protoplanetary nebula to a candidate LBV based on $K$-band spectra and photometry.
Photometric variations of $\approx$$20\%$ are detected in DIRBE data, while mid-infrared imaging confirms the presence of a circumstellar nebula.
The spectroscopic variability exhibited is replicated almost exactly by the bona fide LBV qF362. 
We estimate a distance of 3.68~\kpc, $log(L/{L_\odot})=5.75$, and temperature in the range of 10.5 to 13~kK. 
From the luminosity and temperature, along with a model, we infer an initial mass of $\approx$40~\Msun. 
All of the observed and inferred properties are similar to those of well-established LBVs. 
We consider \iras\ to be a compelling candidate LBV and suggest further photometric and spectroscopic monitoring to confirm this assertion.
 
\begin{acknowledgements}
We acknowledge the efforts to produce the data from IRAS, DIRBE, GSC, USNO, 2MASS, DENIS, GLIMPSE, MIPSGAL, AKARI, and WISE surveys, and the Science Archive of the European Southern Observatory.
We thank John Hillier for providing the CMFGEN code and James Urquhart for information on the radio emission from IRAS~16115$-$5044.
This work was partially supported by the National Natural Science Foundation of China (NSFC- 11421303), and USTC grant KY2030000054. 
F.N. acknowledges financial support through Spanish grants ESP2017-86582-C4-1-R and PID2019-105552RB-C41 (MINECO/MCIU/AEI/FEDER) and from the Spanish State Research Agency (AEI) through the Unidad de Excelencia "María de Maeztu"-Centro de Astrobiología (CSIC-INTA) project No. MDM-2017-0737. 
\end{acknowledgements}

\bibliographystyle{aa}
\bibliography{biblio}

\end{document}